\documentclass{mem}
\usepackage{natbib}\usepackage{txfonts}\usepackage{balance}
\usepackage{graphicx}
\usepackage[a4paper]{hyperref}
\usepackage{epstopdf}

\def\chandra{\emph{Chandra}}

\begin{document}

\title{New High-Resolution Sunyaev-Zel'dovich Observations with GBT+MUSTANG}
\subtitle{}

\author{T. \,Mroczkowski\inst{1,2}
\and M. \,J. \,Devlin\inst{1}
\and S. \,R. \,Dicker\inst{1}
\and P. \,M. \,Korngut\inst{1}
\and B. \,S. \,Mason\inst{3}
\and E. \,D. \,Reese\inst{1}
\and C. \,Sarazin\inst{4}
\and J. \,Sievers\inst{5}
\and M. \,Sun\inst{4}
\and A. \,Young\inst{1}}

\institute{University of Pennsylvania, 209 S. 33rd St., 
Philadelphia, PA 19104.\\
\email{amrocz@sas.upenn.edu}
\and NASA Einstein Postdoctoral Fellow.
\and National Radio Astronomy Observatory, 520 Edgemont Rd., 
Charlottesville VA 22903.
\and Department of Astronomy, University of Virginia, P.O. Box 400325, 
Charlottesville, VA 22901.
\and The Canadian Institute of Theoretical Astrophysics, 60 St.
George Street, Toronto, Ontario M5S 3H8.}

\authorrunning{T. Mroczkowski}
\titlerunning{New MUSTANG High-resolution Sunyaev-Zel'dovich Observations}

\date{Presented 15 November 2010 / Submitted 23 January 2011}
 
\abstract{We present recent high angular resolution (9$''$) Sunyaev-Zel'dovich 
effect (SZE) observations with MUSTANG, a 90-GHz bolometric receiver on the Green 
Bank Telescope. MUSTANG has now imaged several massive clusters of galaxies in 
some of the highest-resolution SZE imaging to date, 
revealing complex pressure substructure within the hot intra-cluster gas in merging
clusters.
We focus on three merging, intermediate redshift clusters here:
MACS~J0744.8+3927, MACS~J0717.5+3745, RX~J1347.5-1145.
In one of these merging clusters, MACS~J0744.8+3927, the MUSTANG observation has 
revealed shocked gas that was previously undetected in X-ray observations.
Our preliminary results for MACS~J0717.5+3745 demonstrate the complementarity 
these observations provide when combined with X-ray observations of the thermal 
emission and radio observations of the non-thermal emission.
And finally, by revisiting RX~J1347.5-1145, we note an interesting correlation
between its radio emission and the SZE data.
While observations of the thermal SZE probe the line of sight integral of 
thermal electron pressure through a cluster, these redshift independent observations
hold great potential for aiding the interpretation of non-thermal 
astrophysics in high-$z$ clusters.}


\maketitle{}

\section{Introduction}\label{sec:intro}

The Sunyaev-Zel'dovich effect (SZE), due to inverse Compton scattering of
photons from the Cosmic Microwave Background (CMB) off electrons in the 
intra-cluster medium (ICM) \citep{zeldovich1969}, has long been sought 
as a probe of cluster astrophysics and cosmology 
\citep[e.g.][]{birkinshaw1991b, carlstrom1996}.
SZE surveys like those with the Atacama Cosmology Telescope \citep{kosowsky2003}, 
the South Pole Telescope \citep{ruhl2004}, and {\em Planck} \citep{rosset2010} 
are now discovering new clusters.
With notable exceptions \citep[e.g.][]{komatsu2001}, these and other SZE 
instruments have resolutions $\sim$1--7$\arcmin$, and therefore do not 
typically probe clusters on scales $\lesssim$~1~Mpc.

Recently, the {\bf MU}ltiplexed {\bf S}QUID/{\bf T}ES {\bf A}rray 
at {\bf N}inety {\bf G}Hz (MUSTANG), a receiver on the 100-m Green Bank Telescope 
(GBT), has observed the SZE from clusters at 9$''$ resolution \citep{mason2010,korngut2011}.
At this resolution, the thermal SZE provides a probe of subcluster scales,
complementary to X-ray studies.
In these proceedings, we discuss these MUSTANG observations and share the preliminary
results from a new observation taken in the 2010C trimester.
We discuss the instrument in \S~\ref{sec:inst},
present recent results in \S~\ref{sec:results}, and discuss 
future directions for MUSTANG in \S~\ref{sec:future}.

\section{GBT+MUSTANG}\label{sec:inst}

MUSTANG is a cryogenic, re-imaging focal plane camera with an $8\times8$ array
of transition edge sensor (TES) bolometers cooled to $\approx 0.3$~K. 
The MUSTANG receiver was built at UPenn and at 90~GHz
is the highest frequency instrument on the GBT.
Using capacitive mesh filters to define the bandpass, MUSTANG has $\approx 19$~GHz
bandwidth for continuum imaging.  
The array has a $0.63 f\!\lambda$ pixel spacing which yields a well-sampled
$42''$ instantaneous field of view with $\sim 9''$ resolution.
For more details, see \citet{dicker2009} and
the MUSTANG web site.\footnote{\tt http://www.gb.nrao.edu/mustang/}

\section{Results}\label{sec:results}

Due to common mode subtraction of the atmosphere, spatial scales larger than $\sim$1.5--2$\times$ 
the $42''$ instantaneous field of view are severely attenuated from MUSTANG observations.
The resulting high-pass filtered observations of the thermal SZE provide a 
measure of line of sight integrated pressure.\footnote{
Other manifestations of the SZE include: the kinematic SZE, 
proportional to the line of sight proper motion of the ICM, the polarized SZE, 
proportional to the transverse motion of the ICM, and the non-thermal SZE, 
proportional to relativistic electrons in the ICM.  See e.g. \citet{carlstrom2002}.}
They are well-situated to complement 
other SZE instruments by measuring cluster substructure on 0.15--1$\arcmin$ scales, 
but do not recover the extended SZE flux used to determine scaling relations 
\citep[e.g.][]{bonamente2008}.
The observations we present here illustrate MUSTANG's utility in 
revealing shock-heated and disturbed ICM.

\begin{figure}[]
\resizebox{\hsize}{!}{\includegraphics[clip=true]{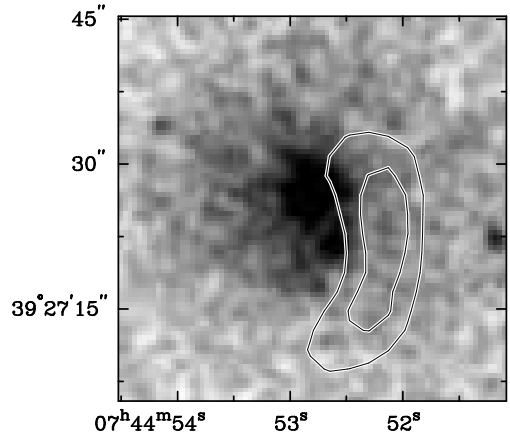}}
\caption{\footnotesize
{\bf MACS~J0744.8+3927:}
\chandra\ X-ray image with MUSTANG SZE 4 and 5-$\sigma$ contours overlaid.
X-ray analysis guided by this SZE feature unveils a significant drop in 
X-ray surface brightness and a rise in temperature, indicating shock-heated gas
with Mach number $\mathcal{M}=1.2 \pm 0.2$.
}
\label{macs0744comp}
\end{figure}

\subsection{MACS~J0744.8+3927 ($z=0.69$)}\label{sec:macs0744}
MACS~J0744.8+3927 was discovered in the ROSAT All-Sky Survey (RASS)
and is part of the {\bf MA}ssive {\bf C}luster {\bf S}urvey \citep[MACS][]{ebeling2001,ebeling2007}, 
a sample of some of the most massive clusters at $z>0.3$. 
As reported in \citet{korngut2011}, MUSTANG observations of this cluster 
revealed a feature in the line of sight pressure, offset from X-ray surface 
brightness peak (see Fig.~\ref{macs0744comp}).  
In fact, the inside edge of the SZE feature aligns with an X-ray surface brightness discontinuity.

Using the MUSTANG observations as a guide, \citet{korngut2011} analysed 
the available 90~ksec of \chandra\ data.  
We interpret the X-ray surface brightness peak as the intact core of the main 
cluster, and the MUSTANG selected feature as shock-heated gas.
The gas to the west of both the MUSTANG selected feature and the second
X-ray surface brightness discontinuity is likely the pre-shock medium.
We derive a best fit Mach number $\mathcal{M}=1.2 \pm 0.2$ \citep{korngut2011},
implying this is a fairly weak shock, marginally consistent with being transonic.

\begin{figure}[]
\resizebox{\hsize}{!}{\includegraphics[clip=true]{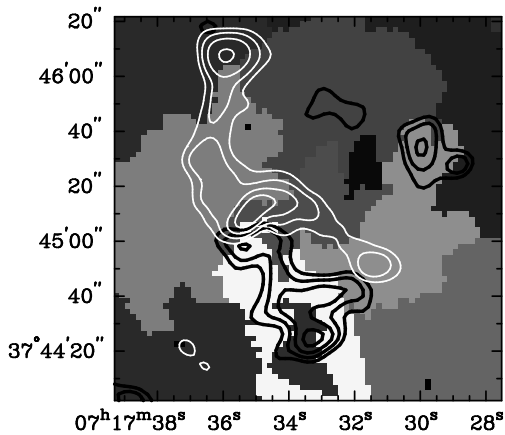}}
\caption{\footnotesize
{\bf MACS~J0717.5+3745 preliminary data:}
\citet{ma2009} temperature map with  MUSTANG $S/N$ (black) and 
GMRT 600~MHz (white) contours from \citet{vanWeeren2009} overlaid.
The peak SZE detection is cospatial with the hottest ($\sim$24~keV, 
in white) region identified in the temperature map,
and is bordered by the radio relic to the north.
}
\label{macs0717}
\end{figure}

\subsection{MACS~J0717.5+3745 ($z=0.55$)}\label{sec:macs0717}
MACS~J0717.5+3745, also found in MACS, exhibits many signposts 
of on-going merger activity.
Optical observations show this cluster to have the largest known
Einstein radius and a complicated mass distribution \citep{zitrin2010}.
Radio observations with the VLA and GMRT have revealed what is likely a
relic \citep{bonafede2009,vanWeeren2009} due to electrons re-accelerated to 
relativistic energies by a shock.
X-ray observations reveal a complex ICM morphology and regions with temperatures
as high as $\sim 24$~keV \citep{ma2009}, indicating a sharp pressure discontinuity
to the south of the radio relic.
The preliminary MUSTANG data help complete this picture: the peak in the MUSTANG
maps agrees well with the position of the hot gas seen in the X-ray (see Fig.~\ref{macs0717}).

\subsection{RX~J1347.5-1145 ($z=0.45$)}\label{sec:rxj1347}
\begin{figure}[]
\resizebox{\hsize}{!}{\includegraphics[clip=true]{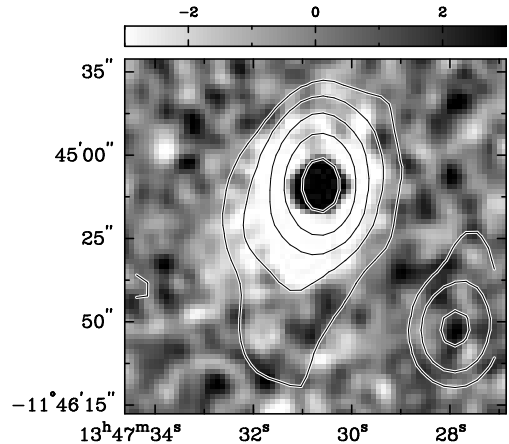}}
\caption{\footnotesize
{\bf RX~J1347.5-1145:}
MUSTANG $S/N$ map with VLA 1.4~GHz contours from \citet{gitti2007} overlaid.
The SZE decrement at $>$3-$\sigma$ corresponds to the radio emission at 
found at $>$9-$\sigma$.  Both are enhanced to the southeast, outside the cool
core. The central point source in the MUSTANG map was detected at $\sim$40-$\sigma$,
and is fully saturated in this map.
}
\label{rxj1347}
\end{figure}

RX~J1347.5-1145, another member of MACS, is the most X-ray luminous cluster known.  While 
its X-ray emission is dominated by its relaxed cool core, there is substantial
evidence pointing to a high impact parameter merger.
High resolution SZE measurements with Nobeyama \citep{komatsu2001} indicated 
(at $\sim$3-$\sigma$) the presence of a pressure enhancement to the southeast.
The Nobeyama result was soon confirmed by X-ray observations \citep{allen2002}.  
MUSTANG later improved upon the significance of the SZE result to many sigma in its first 
cluster observation \citep{mason2010}. 
Interestingly, radio emission from RX~J1347.5-1145, predominantly due to the mini-halo
associated with the relaxed cool core, shows a similar southeast enhancement \citep{gitti2007}.
In Fig.~\ref{rxj1347}, we show that the SZE decrement is co-spatial with the radio emission.

\section{Conclusions \& Future Work}\label{sec:future}

MUSTANG has been pioneering in the study of cluster astrophysics and substructure
through the SZE.  The clusters targeted in MUSTANG observations to date, however, 
have been chosen in an ad hoc fashion.  To address this, we are 
moving to image a cluster sample down to a uniform depth, confirming or rejecting
the presence of cluster substructure that would affect SZE scaling relations 
at the level of a few percent.

We are also working to improve our analysis tools.  Notably, we will soon
be able to remove point source contamination from, and fit cluster models to,
the time ordered data. This will provide a method
by which to fit MUSTANG data jointly with X-ray data and that from other SZE instruments.

Most tantalizing of all, we have proposed a successor instrument to MUSTANG, MUSTANG2,
which is designed to have a larger instantaneous field of view (4.5$\arcmin$) and 
$\gtrsim$~30$\times$ higher sensitivity than MUSTANG.
Observations that currently take hours with MUSTANG could be performed in a matter of minutes,
imaging the SZE at $\sim 9''$ resolution and recovering SZE flux out to $\sim 9\arcmin$ scales.
This upgrade is crucial as the GBT is heavily subscribed, and MUSTANG's high-frequency observations
demand the best observing conditions Green Bank, West Virginia, has to offer.
This upgrade will also ensure GBT+MUSTANG2 will remain competitive as ALMA comes online, 
as MUSTANG2 will provide an order magnitude better mapping speed for continuum emission
at 90~GHz than the 50-element ALMA.

\bibliographystyle{aa}
\bibliography{ntgc}

\begin{acknowledgements} 
We thank M.~Gitti, C.~Ma, and R.~J.~van Weeren for sharing data used for comparison
with the MUSTANG observations presented.
We appreciate the late night assistance from the GBT operators, 
namely Greg Monk, Donna Stricklin, Barry Sharp and Dave Rose.
Support for TM was provided by NASA through the Einstein Fellowship Program, 
grant PF0-110077.
Much of the work presented here was supported by NSF grant AST-0607654.  
Phil Korngut was also funded by the NRAO graduate student support program.
The National Radio Astronomy Observatory is a facility of the National
Science Foundation operated under cooperative agreement by Associated
Universities, Inc. The observations presented here were
obtained with telescope time allocated under NRAO proposal IDs AGBT08A056,
AGBT09A052, AGBT09C059 and AGBT10A056.
\end{acknowledgements}

\end{document}